\documentclass{jpp}
\usepackage{graphicx}
\usepackage{epstopdf, epsfig}
\usepackage{amsmath,amsfonts,amssymb}
\usepackage{subfigure}

\usepackage{xcolor}

\shorttitle{Discontinuous Galerkin Representation of the Maxwell-Jüttner Distribution}
\shortauthor{G. Johnson, A. Hakim and J. Juno}

\title{A Moment-Conserving Discontinuous Galerkin Representation of the Relativistic Maxwellian Distribution}

\author{Grant Johnson\aff{1,2}\corresp{\email{grj@princeton.edu}}, Ammar Hakim\aff{1} and James Juno\aff{1}}

\affiliation{
    \aff{1} Princeton Plasma Physics Laboratory, Princeton, NJ 08540, USA
    \aff{2} Department of Astrophysical Sciences, Princeton University, Princeton, NJ 08544, USA
}

\begin{document}

\makeatletter
\def\fps@table{ht}
\def\fps@figure{ht}
\makeatother

\maketitle

\begin{abstract}
Kinetic simulations of relativistic gases and plasmas are critical for understanding diverse astrophysical and terrestrial systems, but the accurate construction of the relativistic Maxwellian, the Maxwell-Jüttner (MJ) distribution, on a discrete simulation grid is challenging. Difficulties arise from the finite velocity bounds of the domain, which may not capture the entire distribution function, as well as errors introduced by projecting the function onto a discrete grid. Here we present a novel scheme for iteratively correcting the moments of the projected distribution applicable to all grid-based discretizations of the relativistic kinetic equation. In addition, we describe how to compute the needed nonlinear quantities, such as Lorentz boost factors, in a discontinuous Galerkin (DG) scheme through a combination of numerical quadrature and weak operations. The resulting method accurately captures the distribution function and ensures that the moments match the desired values to machine precision.
\end{abstract}


\section{Introduction}\label{introduction}
The Maxwell-Jüttner distribution, originally described by \cite{Juttner1911maxwellsche}, is the relativistic equivalent to the classical Maxwellian distribution. It is the local maximum entropy configuration of a relativistic system of particles and plays the same role in relativistic thermodynamics that the Maxwellian does in non-relativistic thermodynamics. These distributions are central to kinetic modeling, in which the particle distribution function evolves using the Boltzmann, or Vlasov, equation. In these calculations, while the distribution function can be far from local thermodynamic equilibrium, the Maxwell-Jüttner distribution is used for projecting initial conditions, computing approximate collision operators, and for computing differences from local thermodynamic equilibrium for studying and deriving reduced models.

The use of the Maxwell-Jüttner distribution in the particle-in-cell (PIC) framework is well understood and has been detailed, including addressing issues related to particle loading, by \cite{zenitani2015loading}. A growing application of this research is the simulation of high-energy, extreme plasmas, such as relativistic magnetic reconnection (\cite{sironi2014relativistic, guo2014formation}), pulsar magnetospheres (\cite{kuzichev2019magnetospheric, philippov2022pulsar}), and compact objects in curved spacetime (\cite{parfrey2019first, crinquand2020multidimensional, galishnikova2023collisionless}). The recent development of ultra-high intensity laser sources has further motivated new tools for modeling laser-plasma interactions and fundamental instabilities (\cite{derouillat2018smilei, grassi2017electron}). 

Complementary to PIC, kinetic continuum methods are particularly well suited to handle cases where finely detailed resolution in phase space is required to capture turbulence, heat transport, and nonlinearly saturated states--see \cite{juno2020noise} and \cite{nevins2005discrete}. Traditionally, the trade-off for the increased accuracy of continuum methods has been higher computation cost. However, recent developments in improved algorithms such as discontinuous Galerkin (DG) and increases in parallel computing power, particularly with GPUs, have made continuum simulations feasible. Kinetic continuum methods are now seeing newfound broad applicability, for example in codes modeling astrophysical problems such as \texttt{Vlasiator} (\cite{palmroth2018vlasov}), \texttt{Hybrid Vlasov\char`-Maxwell} (\cite{valentini2007hybrid}), \texttt{SpectralPlasmaSolver} (\cite{vencels2016spectralplasmasolver}), and \texttt{Gkeyll} (\cite{juno2018discontinuous}). However, to extend these models to relativistic kinetic systems, we need a Maxwell-Jüttner projection algorithm compatible with high-order schemes and maintains the underlying conservation laws.

The primary challenge in maintaining conservation laws during the projection of the Maxwell-Jüttner distribution onto the finite phase space grid arises from the finite momentum-space extents. These inevitably truncate the distribution function at some finite momentum, thus resulting in a loss of the distribution beyond the maximum momentum represented by the grid. Secondly, the projection onto the grid may alter the moments of the distribution. Without proper moment matching, not only will initial conditions not match the desired macroscopic quantities such as density and temperature, but operators (like reduced collision operators based on the Krook model) will have incorrect moments used to compute them. Finally, the choice of the DG method introduces further complications in determining how to accurately compute nonlinear quantities in the DG basis functions so that we can obtain the desired high-order accuracy from the choice of numerical method.

Here, we develop a novel algorithm that preserves the moments of the distribution and ensures the accuracy of the discretely projected distribution function. While the procedure we discuss is in the context of DG methods, only Sections \ref{sec:DG_specifics} and \ref{subsec:Robust_moments} are DG-specific. The resulting algorithm can be utilized in any continuum kinetic scheme. 

The paper is organized as follows. Section \ref{sec:Moments} describes the Maxwell-Jüttner distribution, the moments of arbitrary relativistic distributions, and how to recover the Maxwell-Jüttner moments from these quantities. Specifically, we show how to arrive at the fluid-stationary frame quantities needed to project the Maxwell-Jüttner from a lab frame simulation. Section \ref{sec:DG} covers the details of the calculations used for DG quantities necessary to project the function and how to initialize the distribution with the proper moments using an iterative scheme. Section \ref{sec:Tests} demonstrates the accuracy of the projection routine, as well as explores an application of it in the relativistic BGK operator. Lastly, we offer concluding remarks in Section \ref{sec:conclusion}.

\section{Relativistic Moments and the Maxwell-Jüttner Distribution}\label{sec:Moments}

The derivation of the Maxwell-Jüttner (MJ) in arbitrary dimensional form is presented in \cite{chacon2010manifestly}. The result, presented in the fluid-stationary frame denoted with primed quantities, is written in the normalized form for a $d$-dimensional momentum space as
\begin{align} \label{MJ}
    f'^{,MJ}(\mathbf{p}') = \frac{n}{2(2\pi)^{\frac{d-1}{2}} (m_0)^d \theta^{\frac{d-1}{2}} K_{\frac{d+1}{2}}(1/\theta)} \exp{\left(\frac{-\gamma(\mathbf{p'})}{\theta}\right)}.
\end{align}
We adopt natural units $c = k = 1$, and define $\theta = T/m_0$, $\gamma(\mathbf{p}') = \sqrt{1+\mathbf{p}'\cdot\mathbf{p}'/m_0^2}$, and $K_{\nu}(1/\theta)$ as the modified Bessel function of the second kind evaluated at $1/\theta$. The distribution is described in terms of the fluid-stationary frame quantities, specifically the number density, $n$, and temperature, $T$. The momentum measured in this, the fluid-stationary frame, is $\mathbf{p}'$. Finally, the particle rest mass, $m_0$, is left unspecified for simulations with multiple species. 

The distribution function, equation (\ref{MJ}), is invariant under Lorentz transformation. This allows us to write the distribution function in the frame observed from the simulation perspective, labeled the lab frame. Variables in the lab frame will be represented with unprimed quantities. Transforming the distribution to the lab frame requires Lorentz transforming unprimed to primed quantities. For the Maxwell-Jüttner distribution, the only quantity that requires transformation is the $\gamma(\mathbf{p}')$ in the exponential. 

Defining the fluid-stationary frame as moving with velocity $\mathbf{v}_b$ relative to the lab frame, the lab frame equation for the MJ distribution becomes
\begin{align} \label{MJ_shifted}
    f^{MJ}(\mathbf{p}) = f'^{,MJ}(\mathbf{p'}) = \frac{n}{2(2\pi)^{\frac{d-1}{2}} (m_0)^d \theta^{\frac{d-1}{2}} K_{\frac{d+1}{2}}(1/\theta)} \exp{\left(\frac{-\Gamma(\gamma(\mathbf{p}) - \mathbf{v}_b\cdot\mathbf{p}/m_0)}{\theta}\right)}.
\end{align}
Here, $n$, and $T$ remain defined as fluid-stationary frame quantities for the density and the temperature. In addition, we define the Lorentz boost factor $\Gamma = 1/\sqrt{1-v_b^2}$, henceforth called the gamma factor, between the fluid-stationary and the lab frame. Note the lab frame is the same as the simulation frame and these terms are often used interchangeably. In the transformation of $\gamma$, 
\begin{align}\label{gamma_trans}
    \gamma' = \Gamma ( \gamma - \mathbf{v}_b \cdot \mathbf{p}/m_0),
\end{align}
we have suppressed the ($\mathbf{p}$) dependence for notational convenience because it is implied. This transform will be utilized again when computing fluid-stationary frame moments, $n$ and $T$, while the distribution function is in the lab frame.

To simplify these equations, we also define the dimensionless quantity, $\mathbf{u} \equiv \gamma \mathbf{v} = \mathbf{p}/m_0$, which replaces all instances of $\mathbf{p}$. We will refer to $\mathbf{u}$ as the momentum, since it is $\mathbf{p}$ divided by a constant rest mass $m_0$. Additional normalization of the spatial lengths, $x$, to a code unit scale so $x$ is also dimensionless, and consequently the distribution function $f(\mathbf{u})$ is dimensionless as well.

The MJ distribution requires fluid stationary moments, $n$,  $\mathbf{v}_b$, and $T$, to project the function onto the grid. These moments can either be specified, such as those given in initial conditions, or computed from an existing arbitrary distribution $f(\mathbf{u})$. In the latter case, we calculate the moments from within the lab frame where the distribution function exists. The next section provides the equations for computing the fluid stationary moments.

\subsection{Moments}\label{sec:moments}
The Maxwell-Jüttner distribution requires the fluid stationary moments $n$, $\mathbf{v}_b$, and $T$ to project the distribution. For operators that depend on the equilibrium distribution, such as the relativistic-BGK collision operator \cite{bhatnagar1954model}
\begin{align}
    \left( \frac{\partial f}{\partial t} \right)_{Coll} = -\nu(f - f^{MJ}), 
\end{align}
the moments of $f^{MJ}$ must match those of the arbitrary distribution $f$ for the operator to remain conservative. Even when these moments are known, such as when projecting a MJ from given initial $n$, $\mathbf{v}_b$ and $T$, moment calculations are still necessary due to differences between the projected distribution and the desired distribution. Therefore, we need to calculate the moments to correct for errors in the discrete distribution. (Section \ref{sec:MJ Init} discusses initialization in more detail.) In this section, we outline a procedure for calculating general moments and how to arrive at the specific quantities needed to project the Maxwell-Jüttner distribution.

The most general form for computing the moments involves the calculation of the flux four-vector and the stress-energy tensor. From these we provide a procedure for both computing the needed MJ moments as well as other diagnostics such as transport coefficients. Measured in the lab frame for an arbitrary distribution, the flux four-vector and stress-energy tensor components are written, respectively, as
\begin{align}\label{eqn:flux_lab}
    F^\mu = \int f(\mathbf{u}) u^\mu \frac{d^3u}{\gamma}
\end{align}
\begin{align}\label{eqn:stress_energy_tensor}
    T^{\mu\nu} = m_0 \int f(\mathbf{u}) u^\mu u^\nu \frac{d^3u}{\gamma}.
\end{align}
Where the four vector $u^\mu$ is defined as $u^\mu \equiv (\gamma, u^i)$. For a comprehensive source on relativistic thermodynamics, see \cite{vereshchagin2017relativistic}.

To compute the boost between the lab and fluid-stationary frames we require the velocity $\mathbf{v}_b$. The velocity can be computed from the lab frame flux four-vector. Suggestively relabeling some quantities of the flux vector makes this more transparent. First, the component $F^{\mu = 0}$ is the density in the lab frame, $N =F^0$. Second, for clarity, the fluxes are relabeled $Nv_b^i = F^i$ for $i = \{1,2,3\}$. It is clear then that the velocity is given by $\mathbf{v}_b = F^i/F^0$.

 The velocity, $\mathbf{v}_b$, gives us all the information we need to transform the lab frame quantities to the fluid-stationary frame. Computing $\Gamma = 1/\sqrt{1-v_b^2}$, where $v_b^2 = \mathbf{v}_b \cdot \mathbf{v}_b$, the components of the Lorentz transform in Cartesian coordinates is given by
\begin{align} \label{lorentz_trans}
  \Lambda_{\phantom{\mu} \mu}^{ \alpha}
  =
  \begin{bmatrix}
     \Gamma & -\Gamma v_{bx} & -\Gamma v_{by} & -\Gamma v_{bz}  \\
     -\Gamma v_{bx} & 1 + (\Gamma - 1)\frac{v_{bx}^2}{v_b^2} & (\Gamma - 1)\frac{v_{bx}v_{by}}{v_b^2} & (\Gamma - 1)\frac{v_{bx}v_{bz}}{v_b^2}  \\
     -\Gamma v_{by} & (\Gamma - 1)\frac{v_{by}v_{bx}}{v_b^2} & 1 + (\Gamma - 1)\frac{v_{by}^2}{v_b^2} & (\Gamma - 1)\frac{v_{by}v_{bz}}{v_b^2}  \\
     -\Gamma v_{bz} & (\Gamma - 1)\frac{v_{bz}v_{bx}}{v_b^2} & (\Gamma - 1)\frac{v_{bz}v_{by}}{v_b^2} & 1 + (\Gamma - 1)\frac{v_{bz}^2}{v_b^2}  
  \end{bmatrix}.
\end{align}
We will call this specific Lorentz transform from the lab to the fluid-stationary frame $\Lambda$. The fluid-stationary frame flux and stress-energy tensor are computed by transforming the lab respective quantities,
\begin{align}\label{eqn:flux}
    F'^\alpha = \Lambda_{\phantom{\mu} \mu}^{ \alpha}F^\mu
\end{align}
\begin{align}
    T'^{\alpha\beta} = \Lambda_{\phantom{\mu} \mu}^{ \alpha}\Lambda_{\phantom{\nu} \nu }^{\beta}T^{\mu\nu}.
\end{align}

The transformed four-vector for the flux, $F'^\alpha = (n,0,0,0)$, directly provides the fluid-stationary frame density $n$. Equipped with $T'^{\alpha\beta}$, $F^\alpha$, and $F'^\alpha$, we can immediately compute the fluid-stationary frame density, velocity, and temperature,
\begin{align} \label{eqn:density_isolate}
    n = F'^0 = N/\Gamma
\end{align}
\begin{align}\label{eqn:vb_isolated}
    v_b^i = F^i/F^0 
\end{align}
\begin{align}\label{eqn:P_isolate}
    P = nT = \frac{1}{3}(T'^{11} + T'^{22} + T'^{33}).
\end{align}
Since we only require the diagonal spatial components, we can simplify the expression for the fluid-stationary frame pressure moment calculation to $\mu = \nu = 1, 2, 3$, which yields,
\begin{align}\label{fluid_stationary_M2}
    P = nT = \frac{m_0}{3} \int f'(\mathbf{u}') u'^2 \frac{d^3u'}{\gamma'}.
\end{align}

We can replace the fluid-stationary quantity, $u'^2$, in the integrand with lab frame quantities, $u'^2 = \gamma'^2 - 1 = \Gamma^2 (\gamma - \mathbf{v}_b \cdot \mathbf{u})^2 - 1$ from equation (\ref{gamma_trans}). In addition, the Lorentz transformation between the primed and unprimed volumes are
\begin{align} \label{volume_transforms_general}
    & \gamma d^3x =  \gamma' d^3x' \\
    & \gamma' d^3u =  \gamma d^3u' \\
    & d^3x\thinspace d^3u = d^3x'\thinspace d^3u'. \label{eqn:invaraint_phase_space_element}
\end{align} 
The last line, equation (\ref{eqn:invaraint_phase_space_element}), is a statement that the phase-space volume element is invariant. These results, pointed out by \cite{zenitani2015loading}, are arrived at by considering proper or canonical intervals of time, length, and momentum. For example, in the case of time, $d\tau$ = $dt/\gamma$ = $dt'/\gamma'$, where $\tau$ is the proper/canonical time interval. Therefore, transforming the volume element with equation (\ref{volume_transforms_general}), and using $f(\mathbf{u}) = f'(\mathbf{u'})$, we have an expression for the fluid-stationary pressure computed with the lab frame quantities
\begin{align}\label{pressure}
    P = nT =  \frac{m_0}{3} \int f(\mathbf{u}) (\Gamma^2 (\gamma - \mathbf{v}_b \cdot \mathbf{u})^2 - 1) \frac{d^3u}{\gamma}.
\end{align}
Dividing out the fluid-stationary frame number density, $n$, leaves us with the temperature, $T$, of the distribution in the fluid-stationary frame. For general d-dimensional momentum space, the $1/3$ coefficient appearing in equations (\ref{eqn:P_isolate}), (\ref{fluid_stationary_M2}) and (\ref{pressure}) becomes $1/d$. In these cases, the integrals and summation are only taken over d-dimensional momentum space.

In summary, the necessary equations to calculate the moments begin with equations (\ref{eqn:flux_lab}) and (\ref{eqn:vb_isolated}) to isolate $\mathbf{v}_b$. Then use $\mathbf{v}_b$ to compute $\Gamma = 1/\sqrt{1 - v_b^2}$ and then to compute $n$ via equation (\ref{eqn:density_isolate}). Finally, $T$ is computed using equation (\ref{pressure}). This provides all the moments of an arbitrary distribution needed to construct a Maxwell-Jüttner distribution with equivalent moments.

\section{Representation of the MJ Distribution}\label{sec:DG}
\subsection{Division and Multiplication in DG, Computing  $\Gamma$}\label{sec:DG_specifics}
The core principle of discontinuous Galerkin schemes lies in representing the evolving quantities as expansions in a chosen basis. For the MJ distribution, we discretize the momentum space into cells and then project the distribution onto a basis expansion within each cell. For further reading on DG schemes, see \cite{cockburn1998runge} and \cite{juno2018discontinuous}. 

The representation of the MJ distribution on a grid requires, at a minimum, discretization in momentum space. In this case the distribution exists at a single point in space and the moment calculations are only functions of time. However, when coupled to advective equations such as Boltzmann or Vlasov equations, we must discretize space as well. With spatial discretization, the moments are now functions of time and discretized in the configuration space. The spatial discretization makes seemingly simple operations, such as the multiplication and division of two DG represented quantities, non-trivial. The result is nonlinear quantities which can not always be computed exactly, discussed later.

DG schemes have two equivalent representations of the local solution. One is a modal representation, in which functions are expanded in terms of a finite basis set within each cell, $\phi_i$. The second is a nodal representation, in which functions are expanded in terms of a series of points within a cell and multiplied by the associated interpolating polynomials. This choice has tradeoffs, but in this section and the next we choose to consistently use a polynomial order $p$ modal representation for the DG scheme. This is because in previous studies we have found that we can utilize modal representations and corresponding weak operations to minimize aliasing errors, which can greatly impact robustness in kinetic calculations by exciting spurious oscillations. See \cite{juno2018discontinuous}, \cite{hakim2020alias} and \cite{hakim2020conservative}. We note that it may be favorable to use nodal representations and tolerate the aliasing errors for these computations because, even if they are more inaccurate, the procedure may be more robust to realizability issues. 

Choosing the polynomial order of the scheme involves a tradeoff. The benefits of higher polynomial orders are increased accuracy, reduced number of required cells, as well as lower numerical diffusion. However, higher polynomial order expansions increase the computational cost and required memory due to the larger number of degrees of freedom in the basis expansion. In higher dimensional simulations, where the phase space can be 4 to 6 dimensional, the number of basis functions grows rapidly. For instance, if a problem has $d$ dimensions and $p + 1$ basis functions per dimension, then the total number of basis functions given by the tensor product is $(p + 1)^d$. In the simulations presented in a later section using the \texttt{Gkeyll} code, we have found from empirical testing that $p = 2$ is the optimal polynomial order for solving the kinetic equation. 

The linear algebra calculations, which arise from weak equality of DG-represented values, can be solved exactly. Take, for example, the calculation of isolating $\mathbf{v}_b$ from $\langle N\mathbf{v}_b \rangle$. Here, the braces, $\langle ... \rangle$, denotes a combined quantity resulting from the moment computation. The expansion of the moments $N, \langle N\mathbf{v}_b \rangle$ in terms of an orthonormal set of basis functions in the configuration space, $\phi_i(\mathbf{x})$, where $\mathbf{x}$ is defined for an interval within a cell $x_i \in [-1,1]$ that has an associated volume, $I$. In each cell, DG quantities are represented as the sum of coefficients times the modal basis
\begin{align} \label{basis_expansion}
\begin{split}
    & N = \sum_i \phi_i(\mathbf{x})\mathbf{c}_{N,i} \\
    & \mathbf{v}_b = \sum_i \phi_i(\mathbf{x})\mathbf{c}_{v_b,i} \\
    & \langle N\mathbf{v}_b \rangle = \sum_i \phi_i(\mathbf{x})\mathbf{c}_{Nv_b,i}. 
\end{split}
\end{align}

The basis functions $\phi_i(\mathbf{x})$ lie in a finite-dimensional function space of order $p$. One such example is the Serendipity basis, presented by \cite{arnold2011serendipity}, which generalizes the Legendre basis to multiple dimensions while removing interior degrees of freedom to reduce the total number of functions required to represent the solution within each cell. While we use the Serendipity basis to represent the solutions in the simulations presented in later sections, the algorithm does not depend on the choice of basis. The only requirement is that an inner product structure exists to isolate coefficients, for example, in computing the coefficients of $\mathbf{v}_b$ from $\langle N\mathbf{v}_b\rangle$ and $N$.

With expressions (\ref{basis_expansion}), we can rewrite $N\mathbf{v}_b = \langle N\mathbf{v}_b \rangle$ in terms of their expansions. Where the left-hand side is two separate quantities $N$ times $\mathbf{v}_b$ and the right-hand side is a single quantity $\langle N\mathbf{v}_b \rangle$. Integrating both sides of this equation gives a relation for the coefficients
\begin{align}\label{weak}
\sum_{i,j} \left(\int_{I}\phi_i\phi_j\phi_kd\mathbf{x}\right)c_{N,i}\mathbf{c}_{v_b,j} = \sum_{i} \left(\int_{I}\phi_i\phi_kd\mathbf{x}\right)\mathbf{c}_{Nv_b,i}.
\end{align}
Evaluating the integrals in parenthesis, the right hand side simplifies to the Kronecker delta, $\delta_{ik}$, for an orthonormal basis. The left hand side integral is a bit more involved, but can be calculated analytically before the simulation begins. The weak equality, equation (\ref{weak}) then can be written as
\begin{align}\label{weak2}
\sum_{i,j} M_{ijk}c_{N,i}\mathbf{c}_{v_b,j} = \mathbf{c}_{Nv_b,k}.
\end{align}
where $M_{ijk} = \int_{I}\phi_i\phi_j\phi_kd\mathbf{x}$ is a 3D array contracted over the $i^{th}$ index with $c_{N,i}$. We can relabel the result as a 2D matrix: $A_{jk} = \sum_{i} M_{ijk}c_{N,i}$. Now this is a clear linear algebra problem. We then find the inversion of $A_{jk}$ to isolate the coefficients $\mathbf{c}_{v_b,j}$ which solves the linear system and isolates $\mathbf{v}_b$
\begin{align}\label{weak3}
\mathbf{v}_b = \sum_{j} \mathbf{c}_{v_b,j}\phi_j = \sum_{j,k} \left( A_{jk}^{-1} \mathbf{c}_{Nv_b,k} \right) \phi_j.
\end{align}
Hypothetically, if we desired to calculate $\mathbf{c}_{Nv_b,k}$ given the other two coefficients, $c_{N,i}$, $\mathbf{c}_{v_b,j}$ we could have simply stopped at equation (\ref{weak2}). 

We now have the tools to solve multiplication and division. Subtraction and addition are trivial operations. However, calculating nonlinear functions, such as $\Gamma = 1/\sqrt{1-v_b^2}$, presents a challenge. We must instead project $\Gamma$ onto the basis and calculate the result with numeric quadrature using the following equation, 
\begin{align}
    c_{\Gamma,i} = \int_I  \frac{1}{\sqrt{1 - \sum_{m} \left(  \mathbf{c}_{v_b,m}\phi_m(x) \right)^2} } \phi_i(x) d\mathbf{x} 
    \approx 
    \sum_{k}  \frac{w_k\phi_i(x_k)}{ \sqrt{1 - \sum_{m} \left( \mathbf{c}_{v_b,m}\phi_m(x_k) \right)^2}  } .
\end{align}
where $w_k$ is the weight of the quadrature at a point $x_k$. Hence $\Gamma =  \sum_{i} c_{\Gamma,i} \phi_i(\mathbf{x})$. 

The approximations in computing the coefficients $c_{\Gamma,i}$ arise from  projecting the nonlinear result back onto the finite polynomial basis, even when the result contains higher order terms.  Operations, such as multiplication, square roots, and exponents of quantities represented on the grid will likewise result in polynomial orders greater than the basis. Projecting the result of these operations back onto the basis gives us a weakly equal projection, where the result and its projection are not pointwise equal, but have matching basis coefficients.

Finally, in implementing this algorithm, we take particular care in the moment computation to ensure the DG scheme is robust and values remain physical. For example, after computing the moment for $\mathbf{v}_b$ and using this quantity to compute gamma. We must ensure that $|\mathbf{v}_b| < 1$ (the speed of light). Further requirements arise particularly from using the velocity moment when converting $\mathbf{v}_b$ to $\mathbf{v}_b^2$ on the spatial grid. In extreme spatial gradient cases, while the velocity is well described, high-order contributions such as $v^2$ to the cell mean can push the mean to be superluminal. Ultimately, to avoid this problem, we retake a lower-order calculation if we find that a modal computation result violates physical bounds on the quantity. A procedure for robustly computing the moments to avoid these issues in the DG representation is presented in the next Section \ref{subsec:Robust_moments}.

\subsection{Robust Algorithm for Computing DG Moments}\label{subsec:Robust_moments}
Numerically, strong gradients in the drift velocity may be super-luminal even if the cell average is less than the speed of light. Constraining the drift velocity to be well-behaved is even harder when trying to compute the DG expanded velocity squared term, $|\mathbf{v}_{b}|^2$, in the Lorentz boost factor and the fluid-stationary frame density. To solve this issue, we applied the following algorithm to the relativistic solver to utilize the spatial component of the bulk four-velocity in the moment calculation. The spatial component of the bulk four-velocity is guaranteed to be well behaved as it does not have to be bounded from above, and computing the needed Lorentz boost factor from this quantity is also guaranteed to be positive definite and greater than one. Thus, we obtain an overall robust scheme for computing moments that will not return unphysical gamma factors. To achieve this refactor the steps are now:
\begin{enumerate}
    \item[(1)] Compute $\mathbf{v}_{b}$ with equation (\ref{eqn:vb_isolated}) using weak division, equation (\ref{weak3}).
    \item[(2)] If $(1 - |\mathbf{v}_{b}|^2) > 0$ at all $p+1$ Gauss-Legendre quadrature points, then 
    \begin{enumerate}
        \item[(a)] Proceed with taking the square root at the Gauss-Legendre quadrature points.
        \item[(b)] Project the result onto the modal basis to construct $\frac{1}{\Gamma} = \sqrt{1 - |\mathbf{v}_{b}|^2}$. For nodal to modal transformations see \cite{hesthaven2007nodal}.
    \end{enumerate}
    \item[(3)] If $(1 - |\mathbf{v}_{b}|^2) < 0$ at any quadrature point, then 
    \begin{enumerate}
        \item[(a)] Evaluate $\mathbf{v}_{b}$ at polynomial order 1 Gauss-Lobatto points. (The Gauss-Lobatto points differ from Gauss-Legendre points in that the Gauss-Lobatto points include the cell vertices. Using them ensures physical values at the boundaries and at all points interior points within the cell by using $p = 1$ representation.)
        \item[(b)] Compute $(1 - |\mathbf{v}_{b}|^2)$ at all the vertices of the cell.
        \item[(c)] If $(1 - |\mathbf{v}_{b}|^2) < 0$ at any Gauss-Lobatto point, floor the value to $1.0 \times 10^{-16}$, which corresponds to a Lorentz boost factor of $1.0 \times 10^{8}$. This is a reasonable upper limit for most astrophysical applications.
        \item[(d)] Reconstruct the modal representation of $(1 - |\mathbf{v}_{b}|^2)$ from these Gauss-Lobatto points using the nodal to modal transformations. 
        \item[(e)] Compute the square root at the peicewise linear Gauss-Legendre quadrature points and project onto the modal basis. This returns a polynomial order 1 expansion of $\frac{1}{\Gamma} = \sqrt{1 - |\mathbf{v}_{b}|^2}$. Note: by using polynomial order 1 and evaluated at the vertices, this projection is guaranteed to be positive everywhere in the interior of the cell.
    \end{enumerate}
    \item[(4)] Compute the fluid-stationary frame density, $n$, using weak multiplication: $n = \frac{N}{\Gamma}$.
    \item[(5)]  Compute the relativistic bulk four-velocity with weak division: $\mathbf{u}_b  = \frac{N\mathbf{v}_b}{n}$.
    \item[(6)] Compute with weak multiplication of $\mathbf{u}_b$ with itself to obtain: $|\mathbf{u}_b|^2$.
    \item[(7)] Compute $\Gamma_\mathbf{u} = \sqrt{1 + |\mathbf{u}_b|^2}$ at Gauss-Legendre quadrature points utilizing the relativistic bulk four-velocity and project the result onto the modal basis. The square root is safely evaluated as $1 + |\mathbf{u}_b|^2$ is strictly positive.
    \item[(8)] Compute the pressure moment in terms of $\Gamma_\mathbf{u}$ and $\mathbf{u}_b$
\begin{align}\label{pressure_modified}
    P = nT =  \frac{m_0}{3} \int f(\mathbf{u}) \left( \Gamma_\mathbf{u}^2 \gamma  -  2\Gamma_\mathbf{u} \mathbf{u}_b \cdot \mathbf{u}+ \frac{( \mathbf{u_b} \cdot \mathbf{u})^2 - 1}{\gamma} \right) d^3u.
\end{align}
\end{enumerate}

\subsection{Maxwell-Jüttner Initialization Routine}\label{sec:MJ Init}
Initialization of the MJ distribution begins with the spatially DG-represented quantities $n, \mathbf{v}_b$, $T$ and $\Gamma$ computed via the routine in Section \ref{subsec:Robust_moments}. These quantities are then used to calculate the MJ distribution, equation (\ref{MJ_shifted}), at the quadrature points. We can then transform the nodal representation onto the modal phase basis, $\phi(\mathbf{x},\mathbf{u})$. 

However, directly evaluating the MJ distribution, equation (\ref{MJ_shifted}), onto the momentum grid has two complications. First, only a limited range of temperatures can be reliably computable because of the exponential behavior of both the distribution's numerator and the modified Bessel function in the denominator.  In particular, the modified Bessel function has significant finite precision errors at low temperatures. Second, in terms of dimension, $d$, the modified Bessel function takes the form $K_{\frac{d+1}{2}}$, which is (potentially) of fractional order. These are notoriously difficult to compute accurately to the desired finite precision.


To avoid the issues of calculating the modified Bessel functions altogether, we replace them with their asymptotic expansions. To leading order, the expansions of $K_1, K_{3/2},$ and $K_2$ are all identical, $K_\nu(x) \sim \sqrt{\pi/2x}\exp{(-x)}$ where $\nu = (1, 3/2, 2)$. They each retain the same leading-order asymptotic behavior as $x \to 0, K_\nu(x) \to \infty$ and as $x \to \infty$, $K_\nu(x) \to 0$. This means that the projected distribution function is offset in magnitude (the density) only. This unnormalized distribution, expanding $K_\nu$ in the Maxwell-Jüttner distribution, equation (\ref{MJ_shifted}), is written as
\begin{align}\label{MJ_unnorm}
    \widetilde{f}^{MJ}(\mathbf{u}) = 
    \frac{n}{2(2\pi)^{\frac{d-1}{2}} (m_0)^d \theta^{\frac{d-1}{2}} \sqrt{\pi\theta/2} }
    \exp\left( \frac{1}{\theta} - \left(\frac{\Gamma(\gamma - \mathbf{v}_b\cdot\mathbf{p}/m_0)}{\theta}\right)\right).
\end{align}
We retain the normalization term in practice because it improves convergence of higher moments. However, it may also be absorbed into the density normalization constant: $\widetilde{f}^{MJ}(\mathbf{u}) = n\exp(...)$. Which is useful when the normalization is too cumbersome to compute or unknown. 

The density moment of equation (\ref{MJ_unnorm}) can be corrected by multiplying the ratio computed via weak division of the correct density moment over the asymptotic density moment to give the properly normalized distribution. Correcting with this density ratio allows us to avoid finite precision errors associated with the modified Bessel functions and simultaneously corrects any errors in the density introduced by the discrete projection. 

There are two common cases we consider for this density correction. The first case arises during initial conditions, when we know the input fluid-stationary frame density and thus can simply use that desired density moment to correct the distribution function and avoid errors in both the modified Bessel functions and projections. The second case is for operators such as the BGK collision operator, where the density is a time-dependent quantity. In this case, we compute the moments described in equations (\ref{eqn:density_isolate}) and (\ref{eqn:vb_isolated}) to obtain the fluid-stationary frame density of the evolving distribution function, and then use that density in the correction routine to ensure that the projected MJ distribution has the exact density of the evolving distribution.

In discrete, finite grids, the higher moments of the MJ projection, $\mathbf{v}_b$ and $T$, may also deviate significantly from the desired values. To reduce the error between the discretely projected distribution's moments, we employ an iterative method. Starting with the desired moments, $n_0$, $\mathbf{v}_{b,0}$, and $T_0$ we create a discrete projection $\tilde{f}$. The moments of this discrete distribution $\tilde{n}$, $\tilde{ \mathbf{v} }_b$, and $\tilde{T}$, will not necessarily match $n_0$, $\mathbf{v}_{b,0}$, and $T_0$. We then use Picard iteration to minimize the difference between these sets of moments. For an iteration $k$, and a vector of the moments $\mathbf{M} = \{n, \mathbf{v}_b, T\}$ we wish to converge, we have an iterations scheme
\begin{align} \label{iterative_correction}
\begin{split}
    & dd\mathbf{M}_k = \mathbf{M}_0 - \mathbf{M}_k  \\
    & d\mathbf{M}_{k+1} = d\mathbf{M}_k + dd\mathbf{M}_k  \\
    & \mathbf{M}_{k+1} = \mathbf{M}_0 + d\mathbf{M}_{k+1}.
\end{split}
\end{align}
Once we have the $k+1$ moments, $\mathbf{M}_{k+1}$, we re-project the distribution function and repeat until the moments converge to a desired error. 

It is important to note that the velocity bounds need to be sufficient for this iteration scheme to converge. For most adequately resolved cases, this algorithm converges to machine precision in between 3 and 20 steps. Significant tail loss of the distribution function on the finite grid slows down the algorithm considerably, as well as having temperatures near the minimum supported temperature for the grid. Taking more than 20 Picard iterations to converge usually means that the distribution function is not realizable for the given moments on the grid. 

An alternative algorithm, presented in Section 3.3 of \cite{dzanic2023positivity}, employs a Newton method which, in some cases, would reduce the number of required iterations. This algorithm is extended to the relativistic Maxwell-Jüttner by replacing the distribution function, using the moments outlined in Section \ref{sec:moments}, and recomputing the partials needed by the Jacobian. However, we found that, for modal DG schemes, the Newton method is prohibitively expensive.  

For the non-relativistic BGK operator, Dzanic, Witherden and Martinelli \cite{dzanic2023positivity} report the Newton correction typically takes 1-2 iterations to reach machine precision. Such an approach could also apply to the Maxwell-Jüttner distribution. However, for specifically the modal DG represented quantities, the Jacobian required by this method would be prohibitively expensive to compute, and each cell would need to invert the Jacobian of the size number of basis, times the number of moments, squared. The trade off is that the corrective method provided here will take additional iterations but is simpler to implement and can automatically handle arbitrary equilibrium functions, such as the MJ, Maxwellian, and equilibrium in curved spaces.

\section{Tests}\label{sec:Tests}
This section explores the accuracy of projecting the MJ distribution function and tests the combination of the projection routine and the moment routine. We address the accuracy of projecting a MJ distribution, examining grid requirements for capturing the distribution within a certain tolerance of error. We demonstrate how the correction routine alters the discrete distribution to match the moments. We also provide analytic estimates for maximum and minimum temperatures representable by the finite momentum grid. Finally, we integrate these concepts into a conservative relativistic BGK operator. All tests in this section were run with the $\texttt{Gkeyll}$ code using polynomial order $p = 2$ Serendipity basis. The scripts are publicly available at: https://github.com/ammarhakim/gkyl-paper-inp. 

\subsection{Convergence Test}\label{subsec:converg_test}
\begin{figure}
\centering
\includegraphics[width = 0.45\linewidth]{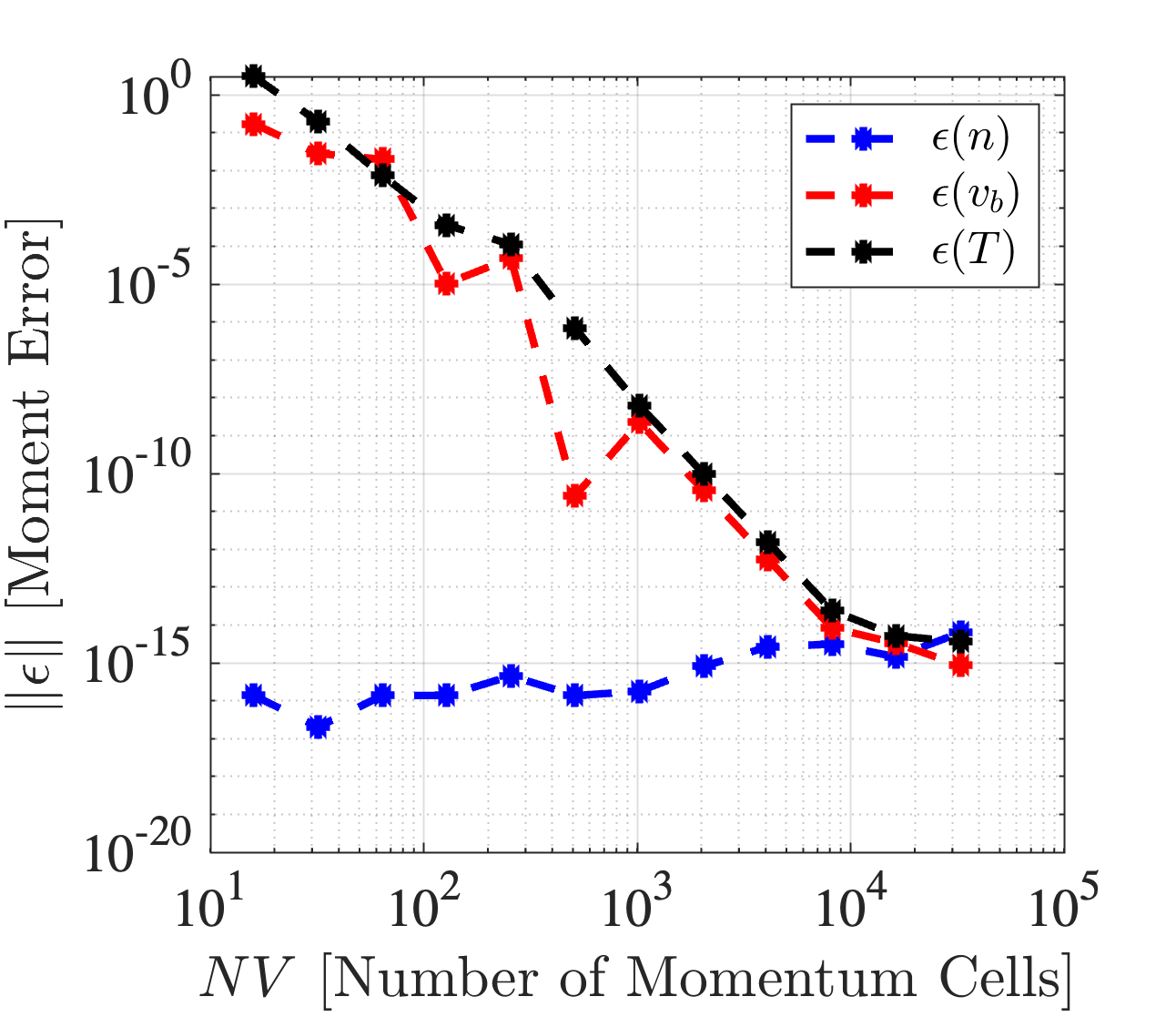}
\caption{Absolute error between the projected Maxwell-Jüttner distribution's moments and the desired moments, without the correction routine for higher moments. The moment values here are $n = 1.0, v_b = 0.5 $, and $ T = 1.0$ and the momentum bounds extend from $u_{max} = \pm 160$. At coarser resolutions, the initial non-monotonicity of the velocity error convergence is caused by small differences in the projection of the distribution onto the discrete grid.}\label{fig:projection_convergence}
\end{figure}

Initially, we project the distribution with only density correction, leaving the higher moments of the projected distribution uncorrected. We then scan the momentum space resolution holding the momentum bounds fixed. We expect, as the distribution becomes better resolved by a finer grid, that the velocity and temperature moments should converge to their exact values. Figure \ref{fig:projection_convergence} illustrates this by plotting the absolute difference in the expected moments and the moments of the projected distribution. As the number of momentum space cells increases, the error approaches machine precision for the moments $\mathbf{v}_b$ and $T$. Hence, with sufficient resolution, we recover the exact velocity and temperature moments, verifying the moments are calculated correctly. As well, the density is normalized properly, as the error remains near machine precision throughout the scan.

We emphasize the reason for machine-matched density is that the density must always be corrected via the rescaling to avoid finite-precision errors in the computation of the modified Bessel functions. We ensure this after each call to create a MJ distribution, the density of the distribution is rescaled to return $n$. Therefore, the density will always agree to the desired value within machine precision. 

From Figure \ref{fig:projection_convergence}, we see that convergence to machine precision without the correction routine of the beam velocity and temperature takes around 10 000 momentum cells with moderate temperature and beam velocity of 1.0 and 0.5, respectively. Convergence indicates we have sufficiently wide momentum bounds to resolve the distribution function for the specified grid size and moments. However, 10 000 grid points is impractical for an actual simulation, especially with higher-dimensional momentum spaces, hence the need for a moment-correction routine. With the correction routine applied to all moments, the same distribution can be represented with only 32 cells, as shown in Figure \ref{fig:BGK}. This reduction and computational savings is one of the central contributions of this paper, making machine precision accurate simulations possible with reasonable domains and with multiple momentum space dimensions. 

The requirement for having a large momentum domain of $u_{max} = \pm 160$ in Figure \ref{fig:projection_convergence} arises from the difference in the MJ distribution's asymptotic behavior for large values of momentum, $u$. Compared with the Maxwellian distribution, this difference becomes apparent. For a 1D Maxwellian distribution, $f^{Max} \sim \exp(-m_0v^2/T)$. In contrast, the 1D Maxwell-Jüttner goes as $f^{MJ} \sim \exp(-m_0|u|/T)$ when $m_0|u|/T \gg 1$. Hence, the Maxwellian approaches zero much faster than the Maxwell-Jüttner on their respective grids. This detail leads to some subtle issues if not minded. To illustrate, the bounds of $u_{max} = \pm 10$ may seem reasonable. However, we are truncating more distribution function than is visually apparent. For example, the lost density fraction for the bounds $u_{max} = \pm 10$ is ($\delta N/N$) $\sim 0.003685$. 

The effect of the missing distribution tails is also apparent when considering the reduced bounds of $u_{max} \pm 10$. In the calculation of the velocity moment, the missing tail changes the expected $v_b = 0.5$ by ($\delta v_b/v_b$) $\sim  0.007342$. This in turn affects the normalization  $n/(N/\Gamma)$ where $\Gamma = 1/\sqrt{1-v_b^2}$ depends on the precision of the velocity calculation. Truncating the tails of the distribution function, especially when the distribution is shifted to having a non-zero bulk velocity, means the velocity moment will always be underestimated. This results in a smaller value for $\Gamma$ and consequently the normalization will produce too small of a density correction. For time-dependent problems, such as the relativistic BGK operator, these errors accumulate. In the BGK test case, without correction of the drift velocity and temperature, the total momentum and energy of the system will decay exponentially.

\begin{figure}
\centering
\includegraphics[width = 0.45\linewidth]{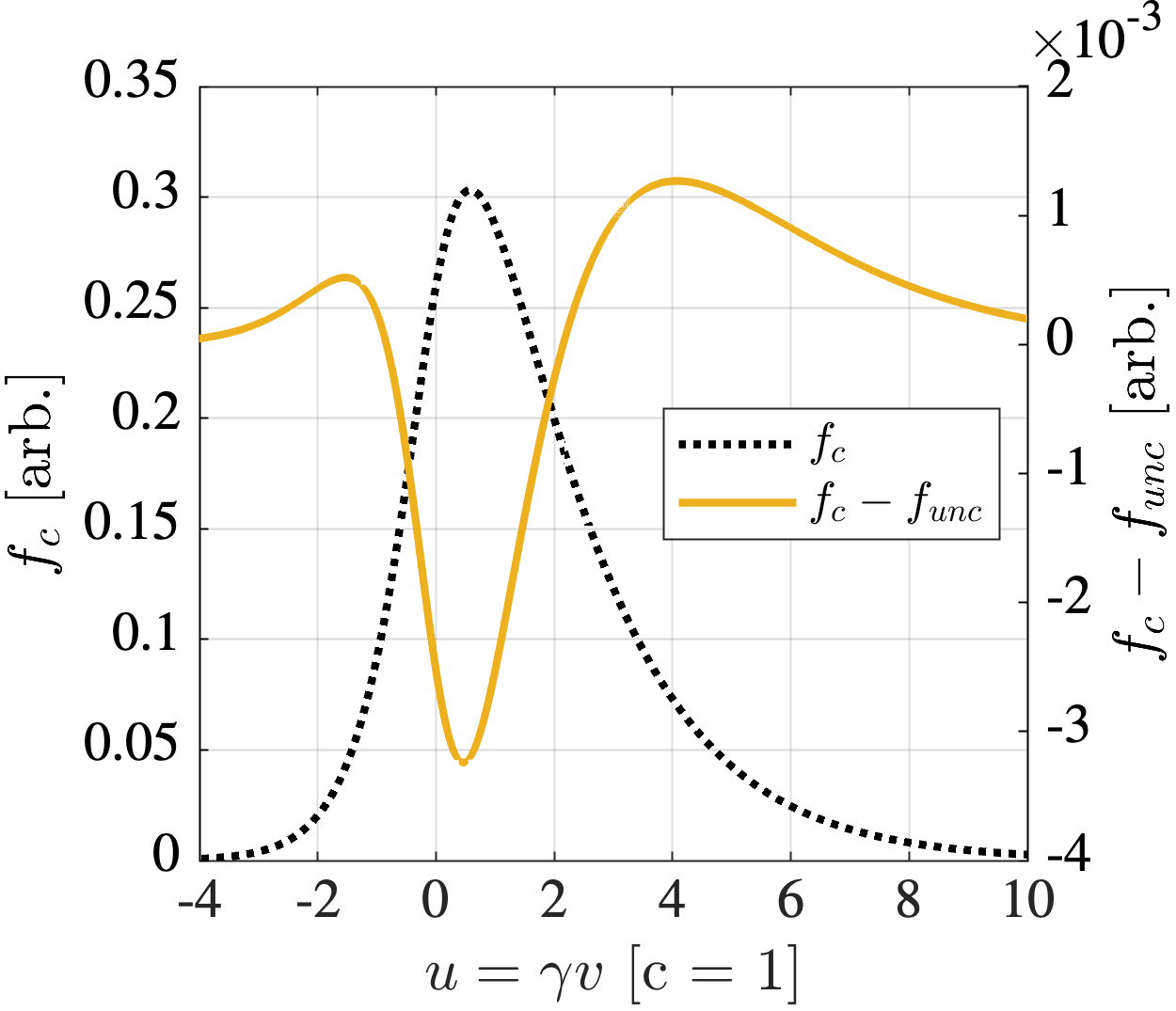}
\caption{MJ plotted with the error between the corrected Maxwell-Jüttner distribution $f_c$ and the uncorrected distribution $f_{unc}$. Since the corrected, uncorrected, and theoretic distributions are indistinguishable on the plot, only the corrected distribution is plotted here.}\label{fig:corr_error}
\end{figure}

If the tails represent a small amount of the distribution function, and the bulk of the distribution is within the grid, then the small errors introduced by the finite grid can be corrected by the iterative scheme from Section \ref{sec:MJ Init}. Fixing the decay issue is our primary motivation for the iterative algorithm to project the MJ. However, it is also useful for accurately creating initial conditions with the proper moments. This does have an effect on the distribution function that is visualized in Figure \ref{fig:corr_error} where we plot a MJ distribution from $u_{max} = \pm 10$ as well as the difference between the corrected and uncorrected distribution functions.

Without the fully resolved right-tail of the distribution function, the iterative scheme creates a MJ with a higher beam-velocity magnitude and temperature to compensate for the losses in the tail. With a higher effective temperature, the peak density is now less pronounced in the corrected distribution and is seen by the negative dip around $u = 0$ in $f_{c} - f_{unc}$. Therefore, even if the projected moments are identical to the desired moments, for insufficiently wide grids, the distribution function is not precisely identical. Under-resolved situations are not explored here because the resulting differences between the exact and projected distribution come down to a failure to represent the function locally with the basis.

\subsection{Representation Limits of Moments of the MJ Distribution}
\begin{figure}
\centering
\includegraphics[width = 0.90\linewidth]{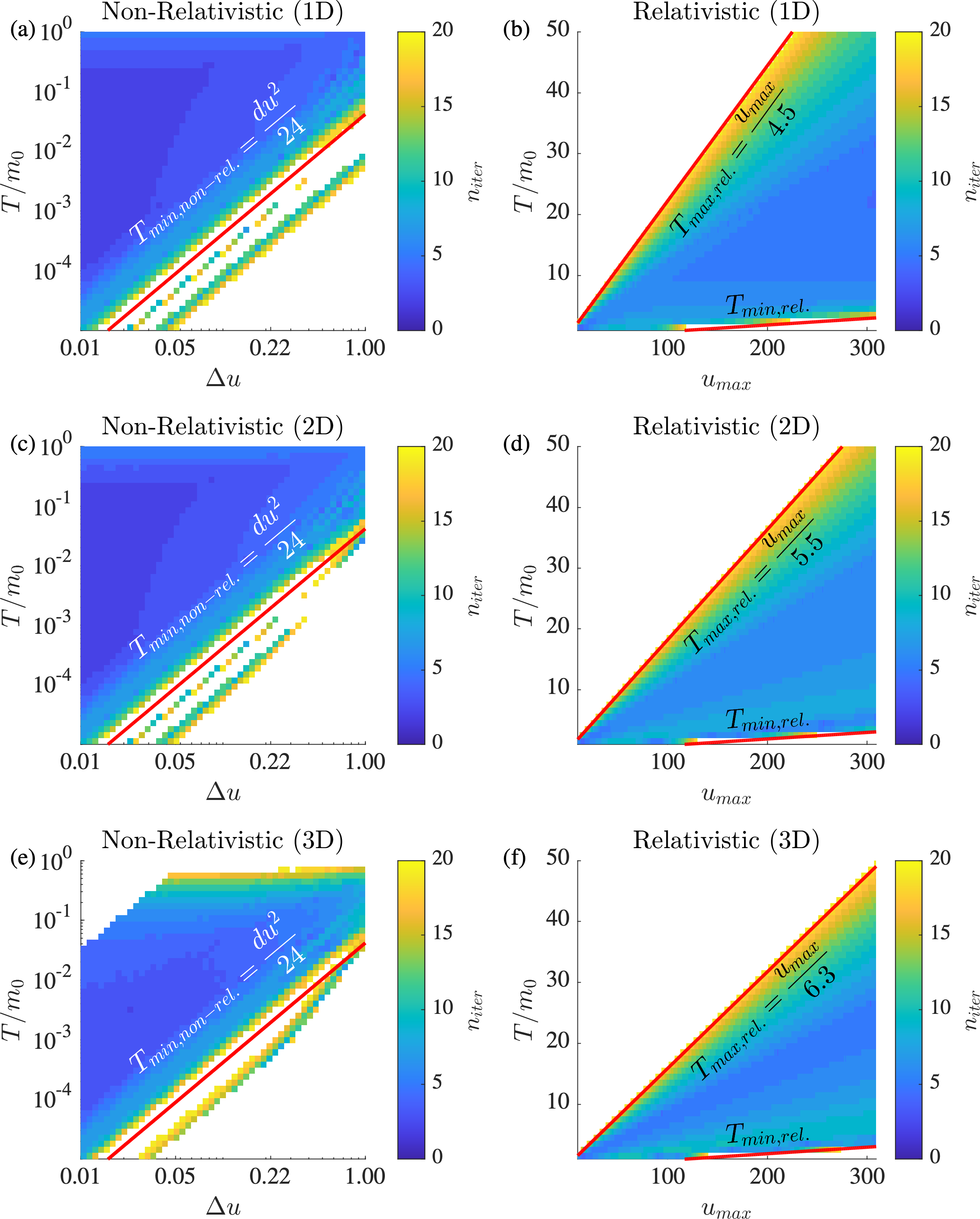}
\caption{Example limitations on the MJ temperature supported by the finite momentum grids. The contours show the number of iterations required for the scheme to converge the moments to an absolute error of $\varepsilon < 10^{-12}$ at varied temperatures and grid parameters. White regions indicate the correction routine took greater than 20 iterations to converge or was unable to converge. Red lines overlay the temperature-limit estimates from this section. Panels (a), (c), and (e) show the non-relativistic limit, while panels (b), (d), and (f) consider relativistic distributions. The rows are ordered by increasing dimensionality of momentum space, from 1D, 2D and 3D.  As a note for panel (e). The upper left corner, colored white, is simulations that were not run due to the large memory requirements. All panels were run with $n = 1$ and $\mathbf{v}_b = 0$. }\label{fig:min_max_temp}
\end{figure}

The discrete grid sets limits on the MJ distribution moments which can be represented. Because the momentum grid only has finite minimum resolution and minimum and maximum momentum bounds, this sets limits on the resolvable momentum and temperature moments of the projected distribution. An immediate restriction is that the momentum moments cannot be beyond the bounds of the domain. The temperature resolution range is more subtle because it affects the width of the distribution. To quantify the resolvable temperatures, Figure \ref{fig:min_max_temp} explores the minimum and maximum temperature allowed in non-relativistic and relativistic limits. 

Figure \ref{fig:min_max_temp} panel (a) scans the case where the bulk of the distribution is non-relativistic for a 1D momentum space. Contours of the number of iterations required to correct the moments in $\Delta u$ vs temperature $T/m_0$ demonstrate that the convergence of temperature is not guaranteed for $T < T_{min, non-rel.}$. Overlaid on the plot, the red line is an estimate of the minimum temperature in the non-relativistic case, which we compute by assuming the distribution is a tent function in a single cell, $f_{tent}(u) = 0.5 - |u|/\Delta u$, and integrating over the momentum bounds, $\pm \Delta u/2$. Substituting this distribution into equation (\ref{eqn:P_isolate}) and taking the non-relativistic limit provides an estimate of the minimum temperature the grid can support. The result is
\begin{align}
    \frac{T_{min, non-rel.}}{m_0} = \frac{\Delta u^2}{24},
\end{align} 
where $\Delta u$ is the grid spacing in momentum space. Depending on the polynomial order of the solution, the shape of the minimum temperature distribution can vary the constant prefactor. Scanning from warmer to cooler temperatures, the number of iterations required to converge the scheme increases. Eventually, the algorithm hits the maximum number of iterations, and occasionally fails to converge below $T_{min, non-rel.}/m_0$. The grid spacing and number of iterations effectively set the minimum grid temperature for the MJ. The trend of minimum temperature holds across 1D, 2D, and 3D momentum spaces.

For warm distributions, the maximum temperature projectable by a MJ on a discrete grid is when the solution is completely flat. The flat distribution sets a maximum temperature supported by the grid of
\begin{align}
    \frac{T_{max}}{m_0} = \frac{u_{max}\sqrt{1+u_{max}^2} - \text{asinh}(u_{max})}{2u_{max}}
\end{align}
which arises from computing the temperature of a flat distribution between finite bounds of equation (\ref{fluid_stationary_M2}) where we have assumed the distribution has no net flow. We also define the inverse hyperbolic trig function as $\text{asinh}(x) = \ln(x + \sqrt{x^2 + 1})$.

In the relativistic limit, Figure \ref{fig:min_max_temp} panel (b), we can simultaneously show the maximum MJ temperature $T_{max, rel.}/m_0$ and the minimum temperature, $T_{min, rel.}/m_0$. Scanning $u_{max}$ and $T/m_0$, a clear region appears where the temperature is able to converge. Like in the non-relativistic case, the maximum temperature projectable by a MJ on this grid is when the solution is completely flat. A flat distribution, when $u_{max} \gg 1$ (highly relativistic) has a maximum temperature $T_{max, rel.}/m_0 \sim u_{max}/2$. However, discretely projecting a flat distribution requires the iterative scheme to drive the input $T/m_0$ to infinity. This takes infinitely many iterations, and since we do not want material at the momentum bounds in any case, we settle for a maximum of 20 iterations which allows a $T_{max}/m_0 \sim u_{max}/4.5$ in this limit. Thus, setting an effective maximum temperature. 

Likewise, for Figure \ref{fig:min_max_temp} panel (b), a minimum temperature can be roughly estimated in the relativistic limit. Assuming a tent function in a single cell, $f_{tent}(u) = 0.5 - |u|/\Delta u$ and integrating between the momentum bounds, $\pm \Delta u/2$ gives us a rough minimum temperature estimate by plugging this distribution into the equation (\ref{eqn:P_isolate}), 
\begin{align}
    \frac{T_{min,rel.}}{m_0} \approx \frac{ (\Delta u^2 + 16)\sqrt{4+\Delta u^2} - 12 \text{asinh}(\frac{\Delta u}{2})\Delta u - 32 }{6\Delta u^2}.
\end{align}
We plot this estimate of $T_{min,rel.}/m_0$ as a red line in Figure \ref{fig:min_max_temp} panel (b).

These limits in both the relativistic and non-relativistic limits hold regardless of dimensionality. Figure \ref{fig:min_max_temp} panels (c) and (d) show the minimum temperature trend in the non-relativistic limit. While in the relativistic limit, in 2D panel (d) and 3D panel (f), the maximum resolvable temperature follows the same trend, but requires additional iterations to converge to the same level of accuracy in projected moments. Meanwhile, the minimum temperature remains unaffected.

A final note for panel (e) when the maximum temperature cannot be resolved. The maximum temperature limits for the non-relativistic cases in panels (a), (c), and (e) are due to the fixed width of the momentum bounds at approximately 10 thermal velocities. These extents become insufficient to resolve the distribution as it approaches $T/m_0 = 1$. Particularity as the dimensionality increased, the relativistic temperatures ($T/m_0 \sim 1$) at higher dimensionality of the grid, the more material of the distribution is left off and the more iterations are required for the iteration scheme to converge. So, the upper temperature limit which appears on this plot is linked to increasing dimensionality, requiring more iterations to converge. This is similar to the upper temperature limit slope weakening with increased dimensionality.

In the white regions of the contours in Figure \ref{fig:min_max_temp}, where the correction routine was unsuccessful, the error between the desired moments and the projected moments begins to rapidly diverge. These areas represent the boundaries of moments resolvable by the grid. Implementing the correction algorithm within the \texttt{Gkeyll} code, we have added checks for convergence between the desired and projected moments at all spatial points and consider the correction routine complete only if all points achieve the desired level of accuracy. If the routine fails, we default to the density-only correction so simulations may continue and have outputs which indicate the failure.

Additional maximum and minimum temperature limits can also be considered. Such as when the distribution shifts too close to the bounds due to strong flows. Generally, good practice is to have checks that allow the simulation to continue but warn the user that the projection routine is failing. This indicates the grid cannot represent the moments with an equivalent MJ distribution.

The correction routine developed in this paper could be complemented with additional techniques such as adaptive mesh refinement (AMR) (for a relevant kinetic example, see \cite{kotipalo2024physics}), non-uniform momentum-space grids, transforming the kinetic equation into the fluid-stationary frame (see \cite{achterberg2018relativistic} and \cite{schween2024using}), or combinations of these techniques. AMR can allow the simulation to flexibly extend its own grid bounds and refine cells if the moments approach the resolvable limits. This would expand the range of the resolvable momentum and temperature moments dynamically, ensuring they successfully converge. Similarly, nonuniform meshes could be employed to coarsely resolve the distribution tail, reducing the truncation error. Lastly, transforming the kinetic equation to evolve in the fluid-stationary frame would eliminate issues with non-symmetric distributions from boosts and eases the grid requirements for challenging configurations such as ultra-relativistic flows. In all of these cases, the correction routine and techniques presented in this paper are still required to maintain conservation laws and enable reasonably compact domains.

\subsection{Relativistic BGK}\label{sec:BGK}
To integrate into a single test the entire projection of the MJ distribution and correction of the moments, we apply these algorithms within a single-species relativistic BGK collision operator. The BGK collision operator will evolve an arbitrary distribution towards a MJ equivalent with matching moments, $n, \mathbf{v}_b$ and $T$. 

The numeric test begins with an initial relativistic shifted water-bag distribution in 1D
\begin{align}\label{waterbag}
f(t=0,u) = f^{WB}(u) = 
\begin{cases} 
      0.5 & 0 < u < 2 \\
      0 & otherwise 
\end{cases}
\end{align}
To robustly test the correction routine, the domain was chosen to span from $u = \pm 4$ and have 32 cells in the momentum space. This represents a grid size used for typical production runs. The representation of the distribution function is polynomial order 2 using the serendipity basis described by \cite{arnold2011serendipity}. The distribution evolves in time via the BGK collision operator given by
\begin{align}\label{eqn:BGK}
\left( \frac{\partial f(t,u)}{\partial t} \right)_{coll} = - \nu \left( f(t,u) - f^{MJ}(u) \right).
\end{align}

\begin{figure}
\centering
\includegraphics[width = 0.45\linewidth]{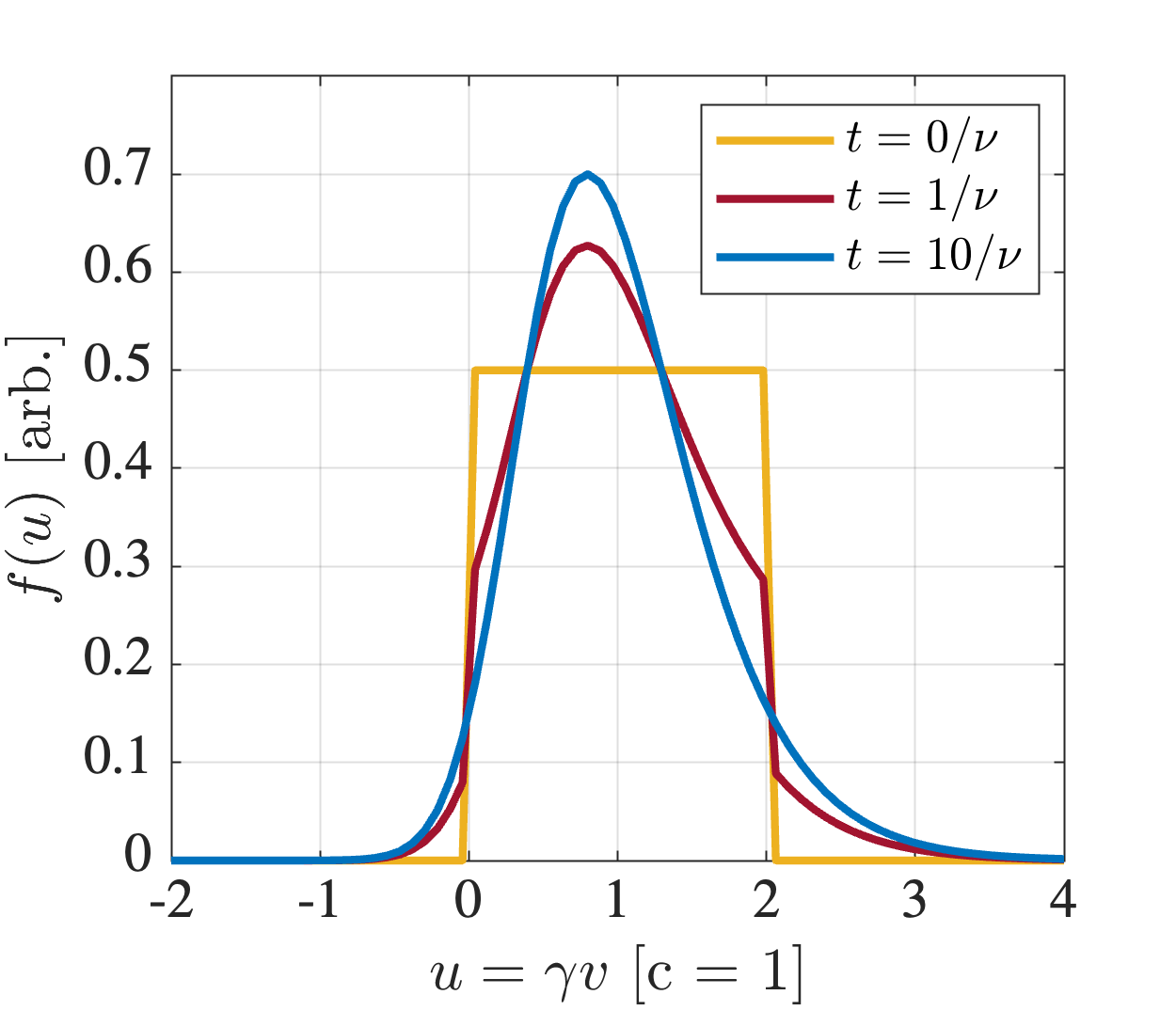}
\caption{Reshaping of the distribution function from a water-bag to Maxwell-Jüttner due by the relativistic BGK operator. The plot includes three time slices: the initial state, one collision-time in, and ten collision-times into the simulation.} \label{fig:BGK}
\end{figure}

Choosing a constant collision frequency of $\nu = 1$, we can exactly solve equation (\ref{eqn:BGK}), $f(t,u) = f^{MJ}(u) + exp(- \nu t) \left( f^{WB}(u) - f^{MJ}(u) \right)$. The solution can be interpreted as the difference between the initial water-bag distribution and the Maxwell-Jüttner distribution, which vanishes the discrepancy with time exponentially until only the Maxwell-Jüttner distribution remains. Figure \ref{fig:BGK} shows the evolution of $f(u)$ at three separate points that match the exact solution of equation (\ref{eqn:BGK}) over time.

Figure \ref{fig:BGK} qualitatively illustrates the expected reshaping of the distribution from a water-bag to a MJ distribution. However, the validation of the scheme comes in the moments of the distribution function. Initially, the water-bag distribution has moments: $n = 0.786$, $v_b = 0.786$, and $T = 0.176$, all with arbitrary units. After ten collision-times, the relative error in each of these quantities with respect to the original quantities are $\epsilon(n) = 1.03\times10^{-14}$, $\epsilon(v_b) = 3.51\times10^{-14}$, and $\epsilon(T) = 2.29\times10^{-13}$. This demonstrates that the moments are conserved throughout the entirety of the BGK collision, maintaining the moments within machine precision of the initial condition.

\section{Conclusion}\label{sec:conclusion}
We have developed and tested a moment-preserving algorithm for projecting the Maxwell-Jüttner distribution onto a discrete momentum grid for application in kinetic continuum simulations. This routine ensures local equivalence of density, velocity, and internal energy between the projected equilibrium and specified moments. This property of the scheme is useful for accurately projecting initial conditions and guarantees conservation laws are maintained by relativistic BGK collisions. When combined with other conservative discrete schemes for the special relativistic Vlasov-Maxwell system, this scheme has the advantage of maintaining conservation laws when extended to the full Vlasov-Maxwell-BGK system.

The distribution and moments are represented and discretized by a discontinuous Galerkin (DG) scheme. The key challenge for DG schemes is computing nonlinear, bounded variables, such as $\Gamma$ or $v^2$ which otherwise can become numerically super-luminal. The scheme we present here robustly handles this challenge by working with the four velocity moments, rather than velocity moments to construct the Maxwell-Jüttner. For added robustness, we set a luminal bound of $\Gamma = 10^8$, which is adequate for most applications. 

Importantly, discrepancies in the moments from finite grid effects are addressed by employing an iterative correction routine to achieve a set precision in the moments of the projected distribution. We performed tests that elucidate how this modifies the projected distribution. Because the correction routine only modifies the input moments of density, velocity, and temperature to the Maxwell-Jüttner, the distribution trades conservation of these lowest three moments for small errors in higher moments. This trade-off is unavoidable as not performing a correction routine leads to artificial decay of all moments due to the finite grid. 

The correction routine further makes possible simulations with reasonable domains and in multiple momentum space dimensions by reducing the grid requirements per dimensions needed to maintain machine precision accurate moments. For a distribution without corrected velocity and temperature moments, Figure \ref{fig:projection_convergence} illustrated convergence to machine precision matched moments takes 10 000 cells and domains of $u_{max} = \pm160$. While Figure \ref{fig:BGK} demonstrated that with the correction routine, a domain of only $u_{max} = \pm 4$ and 32 cells could maintain machine precision in the moments over many iteration of the projection routine.

We additionally demonstrate with these tests the limits of the realizable temperature on the finite grid and provide non-relativistic and relativistic estimations of the lowest and highest temperatures supported on the discrete grid. Beyond these limits, the grid cannot support the distribution and the iterative routine will be unable to converge to a solution. 

We conclude with a test using this routine to relax a waterbag distribution to a Maxwell-Juttner distribution utilizing a relativistic BGK collision operator. After ten collision timescales, the moments of the distribution have remained conserved to about the level of machine precision, thus verifying that the conservation of the lowest three moments holds over successive projections from discrete time stepping.

This work provides a practical algorithm for projecting Maxwell-Jüttner distributions in kinetic continuum simulations of relativistic astrophysical and extreme laboratory settings. This routine can also be leveraged to find closures and transport coefficients by computing perturbations $\delta f = f - f^{MJ}$ to specified accuracy. A combination of ongoing and future work is to extend this algorithm for non-relativistic and relativistic curved-spaces. These extensions immediately work within DG schemes by recycling the algorithm laid out here with the introduction of a spatial metric. Thus, this work has further application enabling collisions for non-Cartesian geometries, which is particularly relevant to astrophysical plasmas around compact objects.

\section*{Acknowledgements}
The authors thank the Gkeyll team, especially Dingyun Liu and Greg Hammett, for useful discussions regarding correction routines.

\section*{Funding}
This work was supported by the U.S. Department of Energy under contract number DE-AC02-09CH11466. G.J. was supported by the U.S. Department of Energy, Office of Science, Office of Advanced Scientific Computing Research, Department of Energy Computational Science Graduate Fellowship under Award Number DE-SC0021110. 

\section*{Declaration of interests}
The authors report no conflict of interest.

\bibliographystyle{jpp}
\bibliography{main}

\end{document}